# Mass hierarchies from two extra dimensions

B. F. Riley

AMEC Nuclear, 601 Faraday Street, Birchwood Park, Warrington WA3 6GN, UK

bernard.riley@amec.com

**Abstract:** The string, MSSM GUT, weak and QCD scales, and the Rydberg constant, correspond to the positions of AdS domain wall intersections in a two-dimensional extra space of infinite extent. The domain walls lie along and parallel to the sides of cells which tile the extra space. The cells are four-sided with parallel opposite sides of lengths $R\ln(\pi/2)$ and $R\ln\pi$, where $R$ is the AdS radius of curvature. Particle masses correspond to the positions of domain walls. We represent the extra space graphically.

## 1 Introduction

In the Randall and Sundrum RS I model [1], an extra spatial dimension is compactified on an $S^1/Z_2$ orbifold and the Standard Model particles and forces are confined to a negative-tension 3-brane, the 'visible' brane, separated from a positive-tension 3-brane, the Planck or 'hidden' brane, by a slice of five-dimensional Anti-de Sitter spacetime ($AdS_5$) of geometry

$$ds^2 = e^{-2k|y|}\eta_{\mu\nu}dx^\mu dx^\nu + dy^2, \qquad (1)$$

where $\eta_{\mu\nu}$ is the Minkowski metric of four-dimensional spacetime, $k$ is the AdS curvature and $y$ is the coordinate of the extra dimension. The exponential factor is the source of the hierarchy between the Planck and weak scales; it results from the small overlap in the fifth dimension of the graviton wave function with our 'visible' 3-brane. In the RS II model [2], the negative tension brane is moved off to infinity; the visible brane is at $y = 0$. Lykken and Randall have combined the results of RS I and II to address the hierarchy problem in an infinite extra dimension with positive tension branes only [3]. Gravity is localized on the Planck brane and we live on a separate brane in the fifth dimension, upon which mass scales are exponentially suppressed. Oda has shown that gravity can be trapped on multiple D3-branes, with positive tension, situated at intervals along the extra dimension [4, 5].

On the basis of the AdS/CFT correspondence [6, 7], Verlinde has observed that the coordinate $y$ in (1) can be thought of as parametrizing the four-dimensional scale [8]: two field excitations at energy scales related by a transformation factor exp($-\lambda$) correspond to two excitations centred on positions related by a translation $y \rightarrow y + \lambda/k$ in the fifth dimension. In Section 2, we will make use of this correspondence between four-dimensional scale and position in an extra dimension to relate particle masses to positions within two extra-dimensional lattices. We will restrict the analysis to quarks, charged leptons, weak gauge bosons, ground state mesons and baryons, atomic nuclei with A ≤ 10, nuclei of double magic number, nuclei in 'islands of stability' and 'α-particle nuclei' with A ≤ 20. In Section 3, we will relate the scales of physics to domain wall intersections in a two-dimensional extra space. In Section 4, we will show that a 'weak' sequence of mass scales corresponds to a lattice that is incorporated within one of the two fundamental lattices.

In Section 5, we will summarize our findings. Values of particle mass have been taken from the listings of the Particle Data Group [9].

## 2 Particle mass scales

We will assume that the higher-dimensional spacetime is AdS, that the Planck scale of the spacetime is equal in size to that of the four-dimensional Planck scale, and that particles occupy lattice points in an extra dimension. A lattice of points extending from the Planck brane with spacing $d/k$ in an extra dimension will correspond to a geometric sequence of mass-energy scales that descends from the Planck Mass $M_P$ (1.220892 x $10^{19}$ GeV) with common ratio exp($-d$). By computation, we have found a geometric sequence of scales, Sequence 1, within which the masses of quarks and baryons are found. The sequence is of common ratio 2/π, to within 1 in $10^5$, corresponding to a lattice spacing of $(1/k)\ln(\pi/2)$. The mass $m_i$ of the $i^{\text{th}}$ level of Sequence 1 is given by

$$m_i = (\pi/2)^{-i} M_{\mathrm{P}}, \tag{2}$$

where $i \geq 0$. Levels are characterized by integer $i$; sub-levels are characterized by half-integer, quarter-integer, eighth-integer, etc, $i$. A particle of mass $m_{particle}$ occupies a level in Sequence 1 for which

$$i = \ln\left(M_P / m_{particle}\right) / \ln\left(\pi / 2\right) \tag{3}$$

Baryons occupy the levels and sub-levels of the mass-energy sequence. The lightest strange, charmed and bottom baryons occupy levels (integer *i*) and 1st order sub-levels (of half-integer *i*), as shown in Figure 1. Horizontal lines in this graph represent both mass-energy levels and domain walls that extend through the lattice points and separate slices of AdS spacetime. Mass-energy sub-levels of $n^{th}$ order correspond to the positions in an extra dimension of domain walls of $n^{th}$ order. Λ (uds, I = 0) shares Level 97 with $\Sigma^0$ (uds, I = 1). $\Lambda_c^+$ (udc) and $\Lambda_b^0$ (udb) occupy 1st order sub-levels. The $I(J^P) = 0(½^+)$ baryon $\Omega_c^0$ (ssc), of mass 2697.5 ± 2.6 MeV [9], occupies a 3rd order sub-level, for which $i$ = 95.125, of mass 2697 MeV. The exotic baryon $\theta^+$, of mass 1534.3 ± 2.5 MeV [9], occupies a 3rd order sub-level, for which $i$ = 96.375, of mass 1533 MeV. Interestingly, the mass difference of Λ and $\Sigma^0$, 76.959 ± 0.023 MeV [9], closely matches the mass, 76.98 MeV, of a level (Level 103). Quarks apparently occupy levels in Sequence 1. The mass of Level 102 is 121 MeV, which is consistent with the strange quark current mass in the $\overline{\mathrm{MS}}$ scheme at a scale μ ≈ 2 GeV [9]. The masses of Levels 97, 94 and 86 are 1.16 GeV, 4.48 GeV and 166 GeV, respectively, which are consistent with the heavy quark running masses in the $\overline{\mathrm{MS}}$ scheme [9, 10].

The lightest charged lepton (electron) and the lightest flavoured mesons ($K^\pm$ and $K^0$) occupy the levels of a second geometric sequence of scales, Sequence 2. The sequence is of common ratio 1/π, corresponding to a lattice spacing of (1/*k*)lnπ. The mass $m_j$ of the $j^{th}$ level of Sequence 2 is given by

$$m_j = \pi^{-j} M_P, \tag{4}$$

where $j \geq 0$. A particle of mass $m_{particle}$ occupies a level in Sequence 2 for which

$$j = \ln(M_P / m_{particle}) / \ln(\pi) \tag{5}$$

The electron occupies Level 45, the K-mesons share Level 39 and the weak gauge bosons $W^\pm$ and $Z^0$ share a 1st order sub-level in Sequence 2, as shown in Figure 2.

Particles apparently occupy levels or sub-levels within both sequences but are usually most closely associated with one or other sequence. The K-mesons occupy a level in Sequence 2 and a 3rd order sub-level in Sequence 1, while Λ and $\Sigma^0$ share a level in Sequence 1 and a 3rd order sub-level in Sequence 2. $W^\pm$ and $Z^0$ share a 1st order sub-level in both sequences. From these observations, it appears that particle masses correspond to positions within a two-dimensional extra space.

## 3 Atomic nuclei

Values of $i$ and $j$ have been calculated for nuclear masses obtained from tables of nuclide mass [11] by subtracting the mass of $Z$ electrons from each value. Values of $i$ range from 85.660 for $^{208}$Pb to 95.929 for $^{2}$H; values of $j$ range from 33.792 for $^{208}$Pb to 37.843 for $^{2}$H. From each number $i$ and $j$, the integer part is subtracted resulting in the fractions $i'$ and $j'$ which may be presented graphically more conveniently than $i$ and $j$. In Figure 3, atomic nuclei, represented by points $(i', j')$, are shown to lie upon or close to fifth and lower order mass sub-levels within Sequences 1 and 2. Included here are the light nuclei $^{2}$H, $^{3}$He, $^{4}$He, $^{5}$He, $^{6}$He, $^{7}$Be, $^{8}$Be, $^{9}$Be and $^{10}$C, the double magic number nuclei $^{16}$O, $^{40}$Ca, $^{48}$Ca, $^{56}$Ni, $^{132}$Sn and $^{208}$Pb, and the $\alpha$-particle nuclei $^{12}$C and $^{20}$Ne. Also included in Figure 3 are $^{54}$Fe and $^{58}$Ni, which fall within the $^{56}$Ni island of stability. Nuclei of even $Z$ tend to occupy the lowest order levels.

Also shown in Figure 3 is the occupation by the pseudoscalar meson isospin multiplets $\pi$, K, D and B of low order mass sub-levels within Sequences 1 and 2. Since the mesons within each multiplet are arranged symmetrically about a sub-level, the geometric mean of the two meson masses is used to calculate $i'$ and $j'$ for the multiplet. The $\phi$ meson and the quarkonium states $\eta_c$, J/$\psi$ and $\Upsilon$ are also included in Figure 3.

The double magic number nucleus $^{40}$Ca occupies a sub-level of fourth order within Sequence 1, as shown in Figure 3. Within the $^{40}$Ca island of stability, the nuclei $^{39}$Ca and $^{41}$Ca occupy sub-levels of seventh order, as shown in Figure 4.

The proton-neutron isospin doublet is arranged symmetrically about a sub-level of seventh order in Sequence 1 and is represented in Figure 5 by a point $(i', j')$ calculated from the geometric mean of the two masses. Also shown in Figure 5 is the $\eta'$ meson, upon a sub-level of seventh order within Sequence 2. The nucleus $^{10}$C also appears in Figure 5. Like many other nuclei, $^{10}$C lies close to a low order sub-level but actually occupies a sub-level of high order.

## 4 The scales of physics

We will conjecture that the two lattices we have identified lie in two different directions in a two-dimensional extra space and that domain walls extending through the lattice points partition the AdS spacetime. The domain walls lie along and parallel to the sides of cells which tile the extra space. The cells are four-sided with parallel opposite sides of lengths (1/k)ln(π/2) and (1/k)lnπ. Four-dimensional scales will correspond to positions in the extra space at equal distances from the Planck brane in the two extra dimensions; that is, with positions on a line equidistant from the two sides of the sector in which $i \geq 0$ and $j \geq 0$. The line is punctuated by domain wall intersections, usually of different order. $0^{th}$ order domain walls intersect on the line at points where $(i/k)\ln(\pi/2) = (j/k)\ln\pi$; that is, where $i/j = \ln\pi/\ln(\pi/2) = 2.535$. This condition is met closely at the points where $i = 33$ and $j = 13$ ($i/j = 2.538$), $i = 71$ and $j = 28$ ($i/j = 2.536$), $i = 109$ and $j = 43$ ($i/j = 2.535$), and $i = 147$ and $j = 58$ ($i/j = 2.534$). The positions of these domain wall intersections correspond to mass-energy scales in a geometric sequence of common ratio $(\pi/2)^{38} \approx \pi^{15}$.

The GUT scale of the MSSM (2 x $10^{16}$ GeV) is shown at the $1^{st}$ order domain wall intersection ($i$, ,$j$) = (14, 5.5) in Figure 6, where one fundamental lattice has been drawn perpendicular to the other. Also included in Figure 6 are the Rydberg constant $R_\infty$ (13.6 eV), the characteristic scale of atomic and molecular physics, at the $2^{nd}$ order intersection (137.5, 54.25), and scales associated with the quarks of each Standard Model family. The bottom and top quarks of Sequence 1 lie at equal distances from the intersection (90, 35.5) in the extra space. The strange and charm quarks of Sequence 1 lie at equal distances from the intersection (99.5, 39.25). The up and down quarks, of mass ~5 MeV [9], lie adjacent to the $0^{th}$ order domain wall intersection (109, 43), of four-dimensional scale 5.12 MeV. In the extra space, the three quark intersections are centred on intersection (99.5, 39.25), which corresponds to a mass-energy of 375 MeV, ≈ QCD scale $\overline{\Lambda}$ [12]. The Higgs field VEV (246 GeV) and the masses of the vector boson $W^{\pm}$ and the vector quarkonia states φ ($s\bar{s}$), J/ψ ($c\bar{c}$) and Y ($b\bar{b}$) also correspond closely to the positions of low order domain wall intersections, as shown in Figure 7. We also note that the sizes of the Planck scale, string scale (5 x $10^{17}$ GeV) and the GUT scale of the MSSM are related: *Planck scale*/*string scale* ≈ *string scale*/*GUT scale* ≈ 25 ≈ $(\pi^{15})^{3/16}$, indicating that the string scale also corresponds to the position of a low order domain wall intersection. Finally, we must mention the $0^{th}$ order domain wall intersection (147, 58), which corresponds to a mass-energy of 0.18 eV and the quasi-degenerate neutrino mass scale [13].

**5 A 'weak' sequence of scales**

Particles of all types considered here occupy the levels and sub-levels of a geometric sequence of scales, of common ratio $(2/\pi)^5$, which descends from the Higgs field VEV and is incorporated within the levels and sub-levels of Sequence 1. The 'weak' sequence corresponds to a lattice of spacing (5/k)ln(π/2) extending from a 'weak' domain wall. The mass $m_w$ of the $w^{th}$ level of the weak sequence is given by

$$m_w = (\pi/2)^{-5w} v, \tag{4}$$

where $v$ is the Higgs field VEV and $w \geq 0$. $W^{\pm}$, hadrons containing strange quarks, and the strange quark of Sequence 1 occupy the levels and lowest order sub-levels of the weak sequence, as shown in Figure 8. $\Omega_c^0$ occupies a level. $W^{\pm}$ occupies a 1st order sub-level. The K-mesons occupy a 2nd order sub-level, as does the exotic baryon $\theta^+$. $\Lambda$ and $\Sigma^0$ share a 3rd order sub-level, as do $D_s^{\pm}$ ($c\bar{s}$ , $\bar{c}s$ ; J = 0) and $D_s{*}^{\pm}$ (J = 1). The strange quark of Sequence 1, of mass 121 MeV, occupies a 3rd order sub-level.

$Z^0$ occupies a 4th order sub-level in the weak sequence. The neutral vector states Y, $B_s{*}^0$ ($s\bar{b}$ ), J/ψ and φ, the muon and the tau lepton occupy levels and sub-levels in a symmetrical arrangement that includes $Z^0$, as shown in Figure 9.

**5 Summary**

First, we showed that quarks and baryons occupy the levels, and sub-levels, of a mass sequence (Sequence 1) that corresponds to a lattice in an extra dimension. Then, we showed that the electron, K-mesons and weak gauge bosons occupy the levels, and sub-levels, of a second mass sequence (Sequence 2) that corresponds to a second lattice. Atomic nuclei occupy the sub-levels of both mass sequences. By conjecturing that the two lattices lie in two directions in a two-dimensional extra space, and that domain walls extend through the lattice points, partitioning an AdS spacetime, we found that particle masses and the various scales of physics correspond to the positions of domain wall intersections in the extra space, at equal distances from the Planck brane in the two extra dimensions. Particles occupy the levels and sub-levels of a 'weak' mass sequence that descends from the Higgs field VEV and is incorporated within Sequence 1.

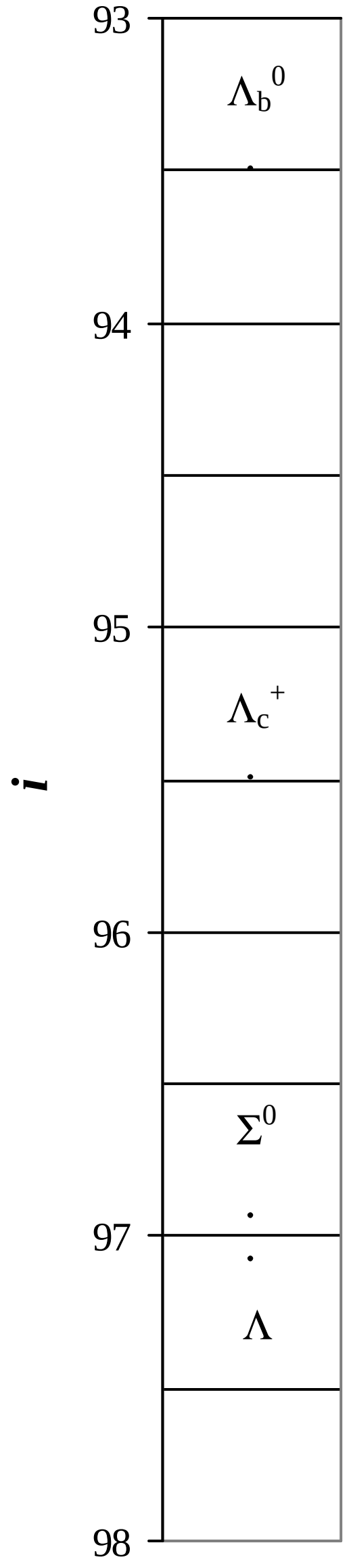

**Fig. 1.** Values of *i* characterizing the lightest strange, charmed and bottom baryons in the mass sequence of common ratio 2/π (Sequence 1).

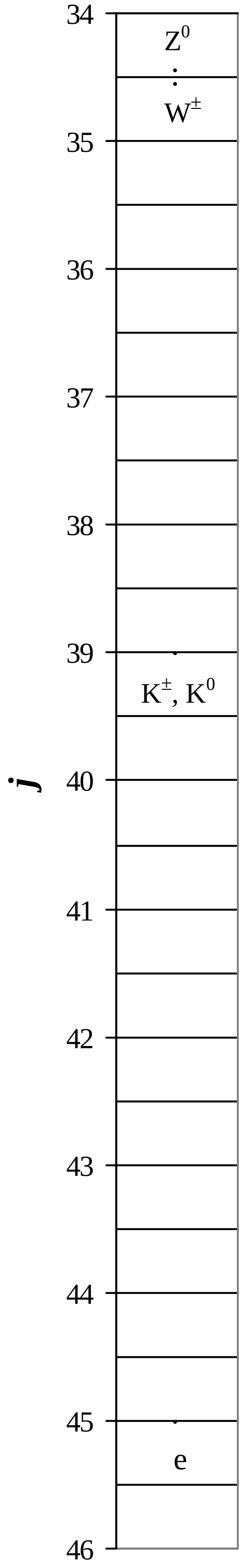

**Fig. 2.** Values of j characterizing the electron, K-mesons and weak gauge bosons in the mass sequence of common ratio $1/\pi$ (Sequence 2).

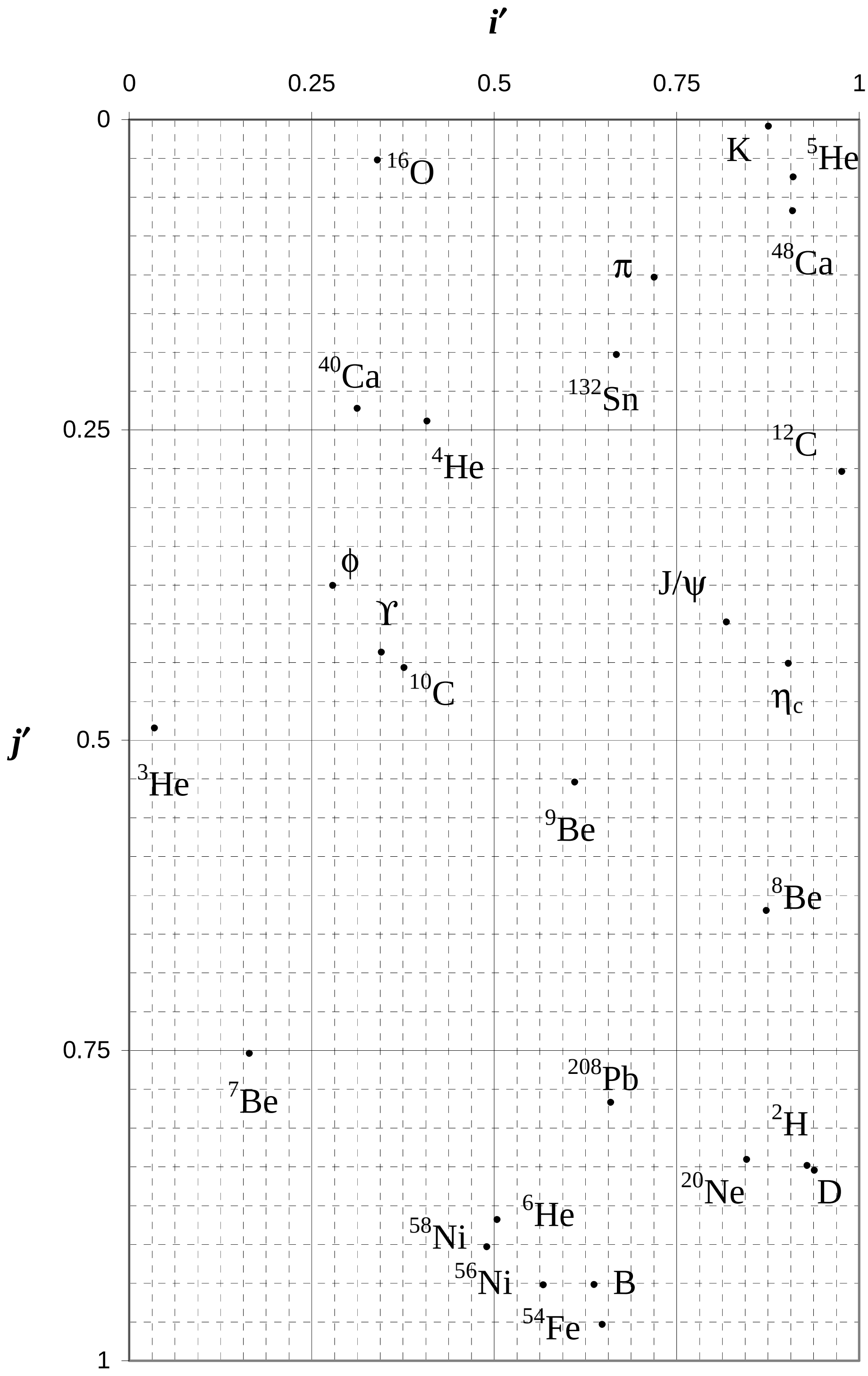

**Fig. 3.** The occupation by mesons and atomic nuclei of mass sub-levels in Sequence 1 and Sequence 2. π, K, D and B refer to pseudoscalar meson isospin multiplets, each of which is represented by the geometric mean of the two particle masses.

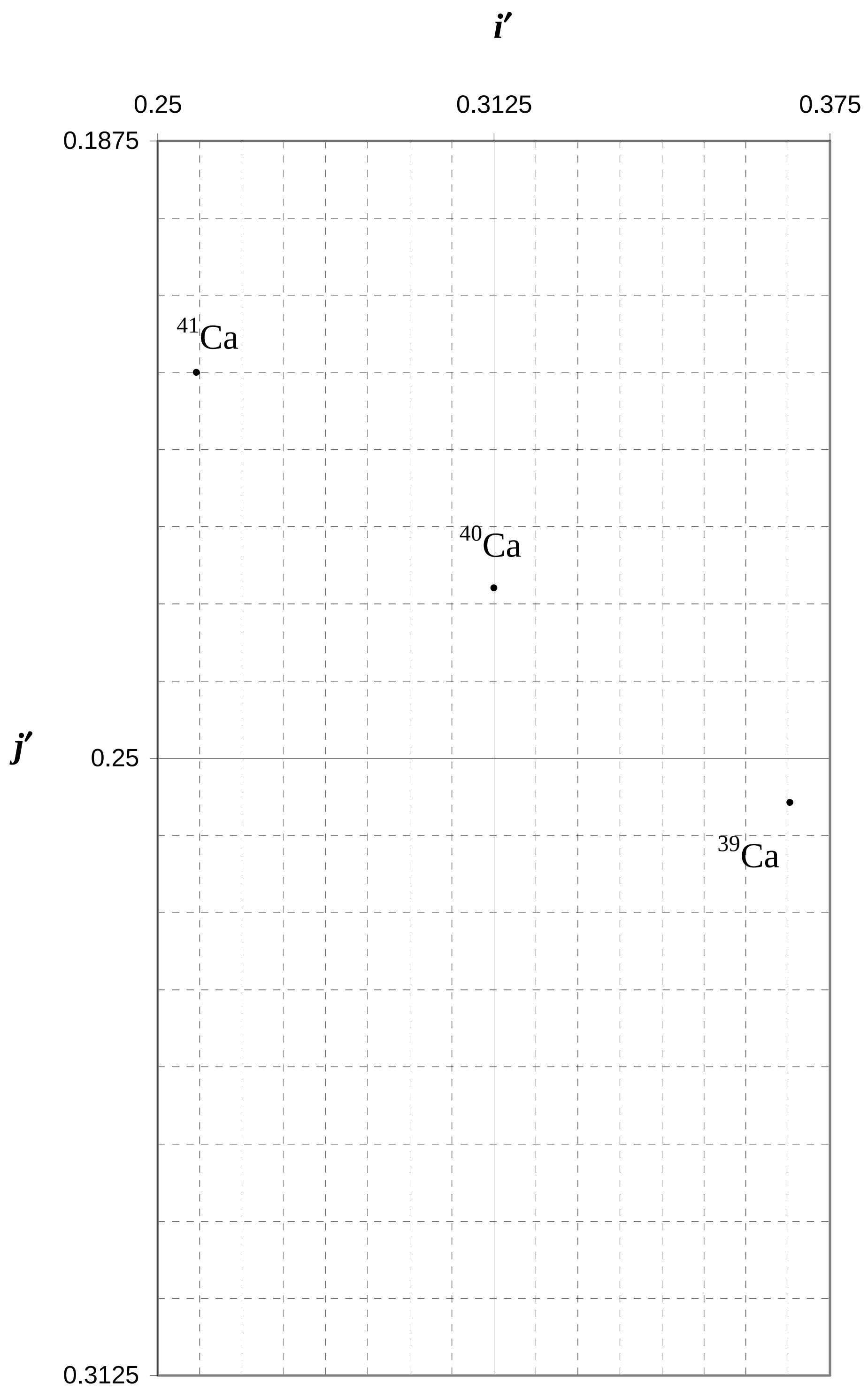

**Fig. 4.** The occupation by calcium nuclei of mass sub-levels in Sequence 1 and Sequence 2

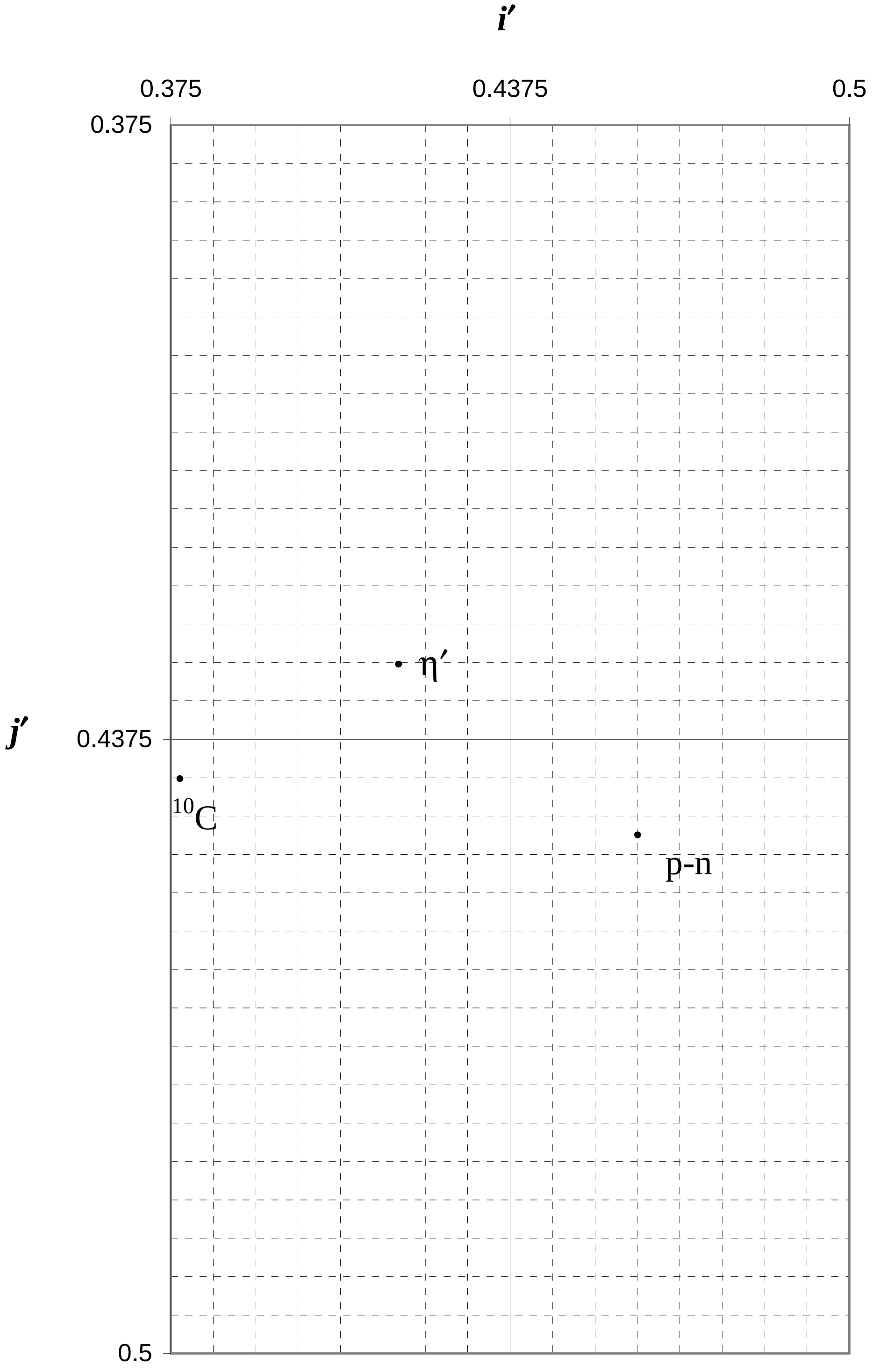

**Fig. 5.** The occupation by assorted particles of mass sublevels in Sequence 1 and Sequence 2. The proton-neutron isospin multiplet is represented by the geometric mean of the two particle masses.

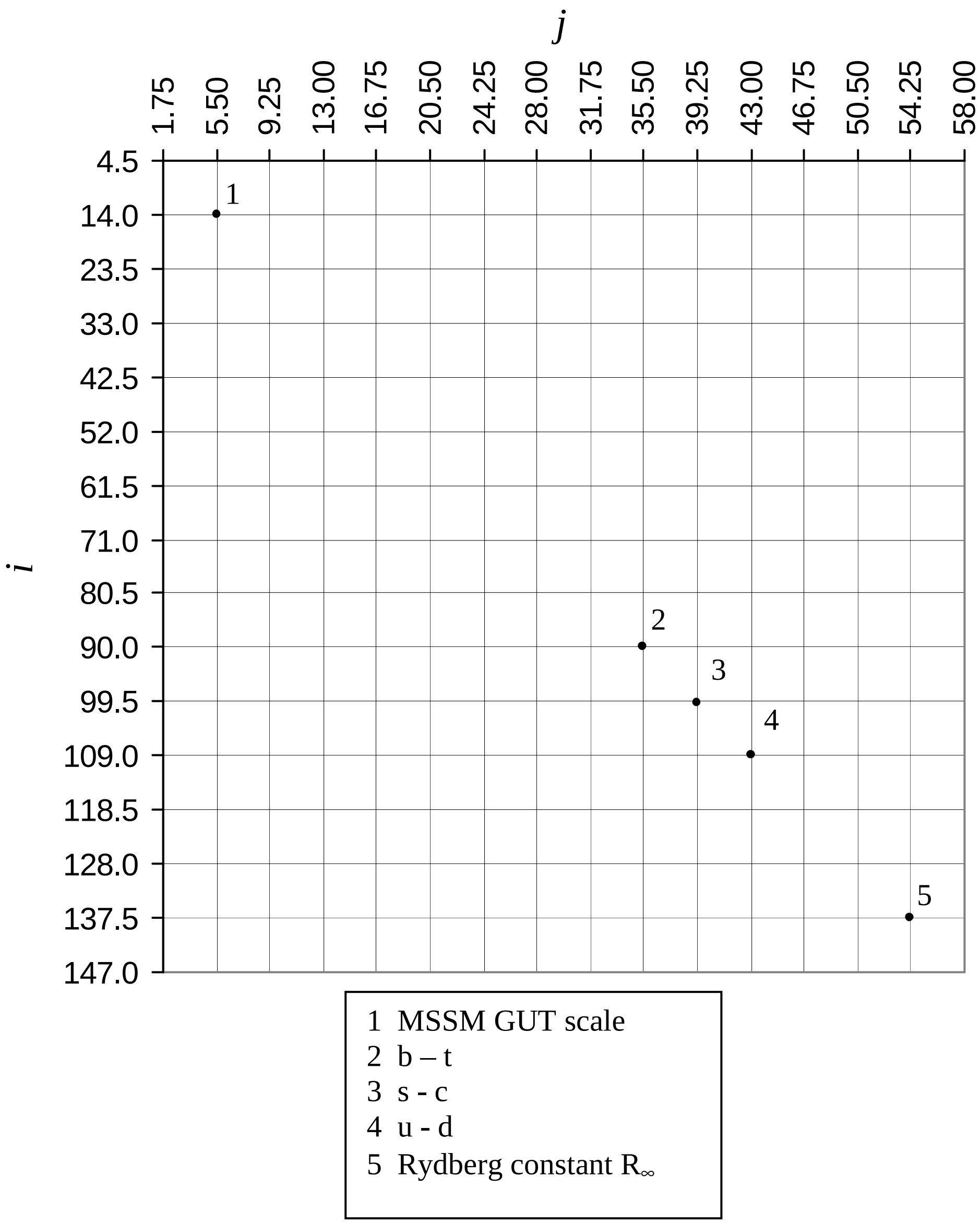

**Fig. 6.** Domain wall intersections in the extra space and the corresponding positions of four-dimensional scales, 1.

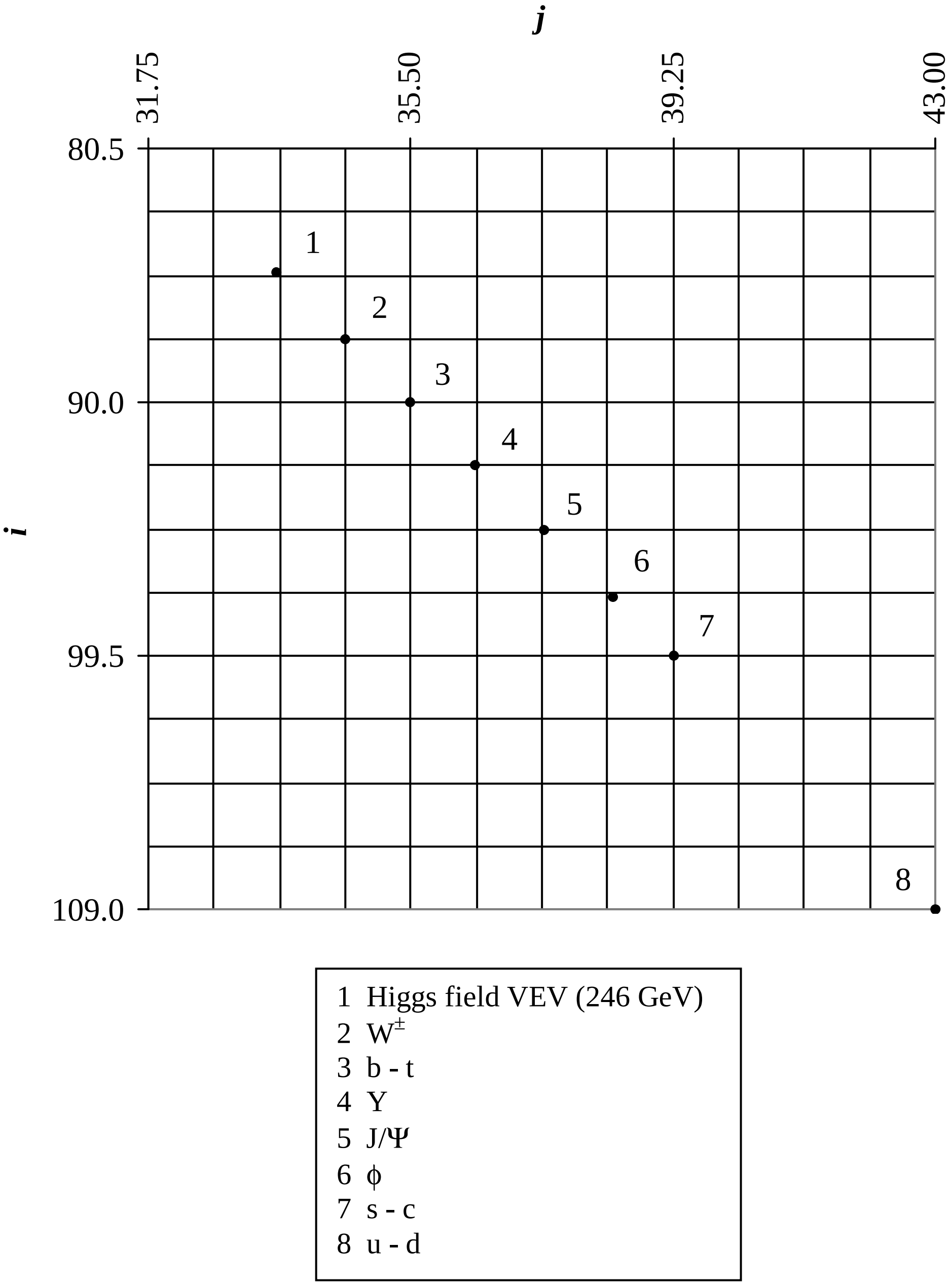

**Fig. 7.** Domain wall intersections in the extra space and the corresponding positions of four-dimensional scales, 2.

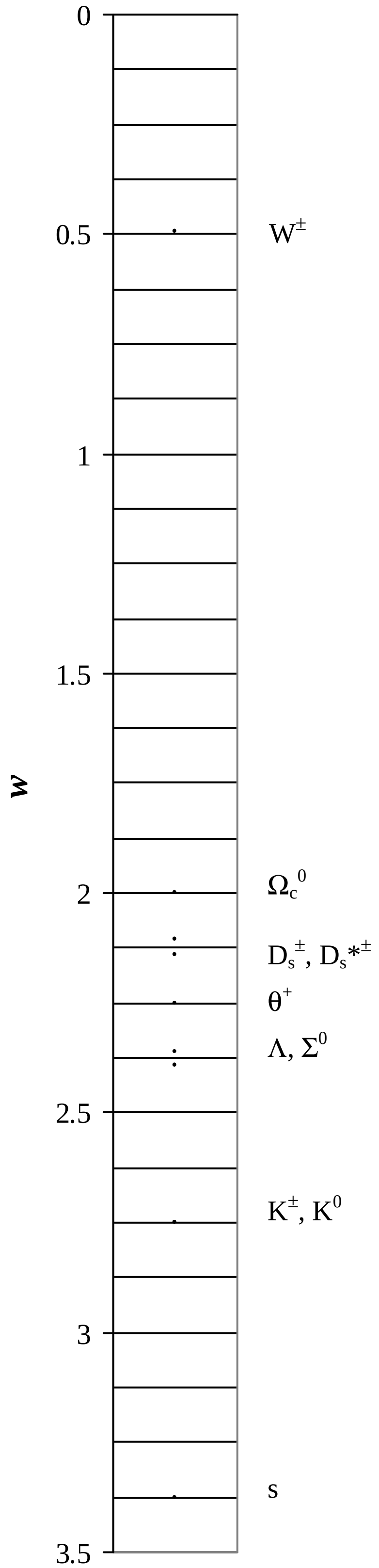

**Fig. 8.** Values of *w* characterizing particles in the weak mass sequence of common ratio $(2/\pi)^5$.

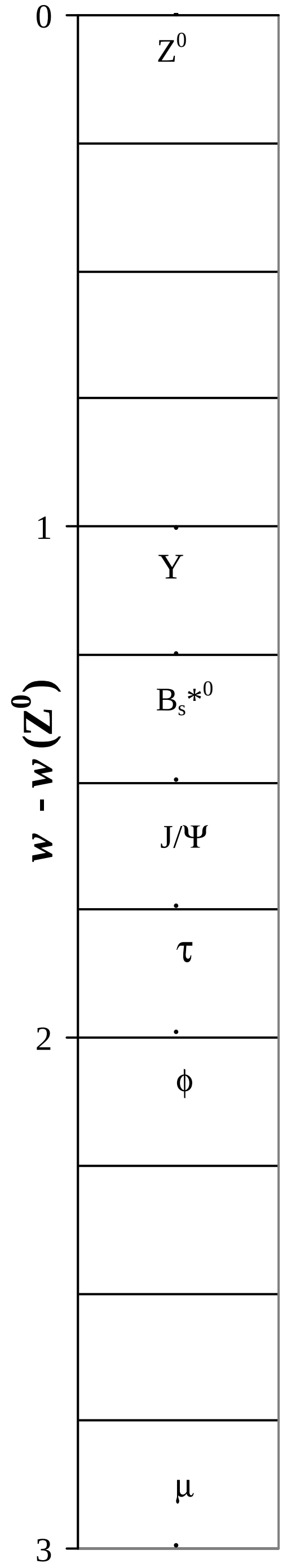

**Fig. 9.** Values of *w* relative to *w*($Z^0$) characterizing neutral vector mesons and charged leptons in the weak mass sequence of common ratio $(2/\pi)^5$.